\documentclass{pasj00}
\draft

\begin{document}
\SetRunningHead{Y. Itoh}{Tidal mechanism}

\title{Tidal mechanism as an impossible cause of the observed secular 
increase of the astronomical unit}

\author{Yousuke \textsc{Itoh}}
\affil{Astronomical Institute, Graduate School of Science, Tohoku
  University, Sendai Miyagi 980-8578, Japan}
\email{yousuke@astr.tohoku.ac.jp}

%


\maketitle

\begin{abstract}
\citet{KB2004} reported a secular increase of the astronomical unit (AU) of 
$15$ meters per century.
Recently, \citet{Miura2009} proposed that a possible angular momentum 
transfer from the rotation of the Sun to 
the orbital motion of the solar system planets 
may explain the observed increase of the AU. 
They assumed that the tidal effect between the 
planets and the Sun is the cause of this transfer. 
Here we claim that tidal effect cannot be a cause of this type of
the transfer to explain the increase of the AU.
\end{abstract}

It is well known that the mean separation between the Earth and the 
Moon has been increasing due to the tidal effect. The gravitational
force  of the Moon induces tidal deformation on the Earth and this deformation 
then exerts a tidal torque on the Moon. Then the orbital angular 
momentum of the Moon increases while the Earth loses the rotational 
one, thereby the orbital radius of the Earth-Moon system gets longer.   

Recently, \citet{Miura2009} proposed that this scenario may give 
an explanation on the secular increase of the astronomical unit (AU) 
of 15 meters per century reported by \citet{KB2004}. 
Namely, they assumed some tidal mechanism that transfers the 
angular momentum from the Sun rotation to the planets orbital motions and 
evaluated the necessary amount of the angular momentum transfer. They
found that the decrease of the Sun's rotational period is too tiny to 
be observed. 

In this paper we explicitly evaluate the increase of the planet orbital 
radius $a_p$ due to the tidal effect.  The theory of the 
secular increase of a satellite orbital radius 
around a planet is well known and given by, for example,  
Eq . (4.160) of \citet{MD1999}. Applying it to a planet-Sun sytem, the
secular increase in $a_p$ becomes  
\begin{eqnarray}
\dot a_{p} &=& 
3\kappa_{2\odot}
\frac{m_p}{m_{\odot}}
\left(\frac{C_{\odot}}{a_p}\right)^5 n_p a_p
\sin2\epsilon_{\odot},
\label{eq:adot} 
\end{eqnarray}
where a circular orbit is assumed to obtain an order of magnitude estimate.
The subscript ``p'' denotes a planet that induces the 
hypothetical tidal effect on the Sun and runs from the Mercury, 
the Venus, and so on. 
$m_p$ and $m_{\odot}$ are the mass of the planet and 
the Sun, respectively.
$n_p$ is the orbital mean motion of the planet and the over-dot 
denotes a time derivative. 
$C_{\odot}$ is the mean radius of the Sun.
$\epsilon_{\odot}$ is the tidal lag angle. 
$\kappa_{2\odot}$ is the tidal Love number of the Sun. 
The tidal Love number of the Sun itself may be 
difficult to measure, but can be estimated from 
the theory and the observation of apsidal motions of 
stars in eclipsing binaries.
Both the theory and the observation shows $\kappa_2$ 
for a solar mass main sequence star having the solar metallicity 
is $\log \kappa_2 \sim - 1.1$
(See e.g., Sec. 18 of \citet{Schwarzschild1965} and 
Sec. 7.2 and Fig. 12 (a) of \citet{Torres2009}. 
Note that the tidal Love number $\kappa_2$ is 3 times the apsidal motion
constant denoted as $k$ in \citet{Schwarzschild1965} 
or $k_2$ in \citet{Torres2009}).

The table \ref{tab:adot} shows the values of 
$\dot a_{p}$ for the 5 inner-most planets. For all of those planets,  
$\dot a_{p}$ is well below the reported value of the increase of the AU.
Note that the value of $\dot a_p$ for the Earth is about 100 times larger than 
that for the Mars. Note also that the most accurate observational 
data for the planetary motion is from the Earth-Mars distance 
measurement (e.g., \citet{Standish2005,Pitjeva2005}). 
So let us consider the Earth-Mars distance. Then, whatever 
large $\kappa_{2\odot}\sin2\epsilon_{\odot}$ is, one should in principle see that the distance 
between the Mars and the Earth in opposition to the Sun 
secularly decreases and that in conjunction secularly increases. 
This is not the same as the increase of the AU, as it should cause  
homogeneous expansion of the planetary orbits. 
After all, 
it is clear from the orbital radius dependence in Eq. (\ref{eq:adot})  
that 
the tidal interaction cannot cause homogeneous expansion of the planetary
orbits.
\begin{table}[ht]
  \caption{
The estimated increase rates in the orbital radii 
for the 5 inner-most solar planets  
due to the Sun's tidal torque acting on those planets. 
Eq. (\ref{eq:adot}) is used for those estimates. The values of   
 $(d a_{p}/dt)/(\kappa_{2\odot}\sin2\epsilon_{\odot})$ 
in the unit of millimeters/century
are shown.}\label{tab:adot}
  \begin{center}
    \begin{tabular}{ccccc}
      \hline
      Mercury & Venus & Earth & Mars & Jupiter  \\
      $6.6$ & $3.1$ & $0.65$ &
		 $0.0068$ & $0.024$ \\
      \hline
    \end{tabular}
  \end{center}
\end{table}

As the AU is such a fundamental quantity in the solar system 
planetary physics (e.g., \citet{Standish2005,Capitaine2008}), 
several works have attempted to explain the increase of the AU. 
However, (1) the real error of 
the value of the AU itself is 3 meters \citep{PS2009}, 
(2) from the measurement of the planetary mean motion, the increase 
of the AU may be less than 5 meters per century \citep{KB2004}  
and (3) it seems that recent studies give smaller values for the 
increase of the AU ($\sim$ 1 meter/century. See \citet{Noerdlinger2008}). 
Hence the increase of 
the AU does not seem robust. On the other hand, radiative mass loss 
of the Sun would increase the planetary orbital radius 
of a meter level over a century \citep{Noerdlinger2008}, and 
we expect to hear in the near future a result of such a fundamental 
test of physics, the equivalence of the radiative energy and the 
gravitational mass.

\section*{Acknowledgments}
Prof. Umin Lee  at Tohoku University let the author know 
the reference on the tidal Love number, for which the author 
is grateful to him.  
I thank the referee for carefully read this paper, corrected the errors 
I made and kindly gave 
me useful comments that have significantly improved this paper. 
The author is supported by 
the Japan Society of the
Promotion of Science (JSPS) Global Center of Excellence (COE) Program (G01): 
Weaving Science Web beyond Particle-Matter Hierarchy at Tohoku
University, Japan.


\end{document}